\documentclass[preprint,12pt, a4paper]{elsarticle}

\usepackage{amssymb}

\usepackage{lineno}

\usepackage{xcolor}

\usepackage[newfloat, draft]{minted} 
\newcommand{\python}[1]{\mintinline{python}{#1}}
\setminted[python]{ %
    linenos=true,             
    autogobble=true,          
    frame=lines,
    framesep=2mm,
    fontsize=\footnotesize
}
\newcommand{\cinline}[1]{\mintinline{c}{#1}}
\SetupFloatingEnvironment{listing}{placement=htp}
\usepackage{caption}

\usepackage[english]{babel} 
\usepackage{tikz-uml}

\usepackage{lscape}

\usepackage{hyperref}

\journal{SoftwareX}

\begin{document}

\begin{frontmatter}



\title{COFFEE - An MPI-parallelized Python package for the numerical evolution of differential equations}


\author[georgead]{Georgios Doulis}
\author[joergad]{J\"{o}rg Frauendiener}
\author[chrisad]{Chris Stevens}
\author[benad]{Ben Whale}

\address[georgead]{Institute of Theoretical Physics, Department of Physics, University of
Warsaw, Warszawa, Poland}
\address[joergad]{Department of Mathematics and Statistics, University of Otago, Dunedin 9054, New Zealand}
\address[chrisad]{Department of Mathematics, Rhodes University, Grahamstown 6140, South Africa}
\address[benad]{School of Mathematics and Applied Statistics, University of Wollongong, Wollongong, NSW 2522, Australia}

\begin{abstract}
COFFEE (Conformal Field Equation Evolver) is a Python package primarily
developed to numerically evolve systems of partial differential equations over
time using the method of lines. It includes a variety of time integrators and
finite differencing stencils with the summation-by-parts property, as well as
pseudo-spectral functionality for angular derivatives of spin-weighted
functions. Some additional capabilities include being MPI-parallelisable on a
variety of different geometries, HDF data output and post processing scripts to
visualize data, and an actions class that allows users to create code for analysis
after each timestep.

\end{abstract}

\begin{keyword}
Python \sep Differential equation solver \sep Parallelized



\end{keyword}

\end{frontmatter}




\section{Motivation and significance}
  \label{sec:motandsign}







  We present a software package, 
  the Conformal Field Equation Evolver or COFFEE for short, that implements
  techniques suitable for numerical solution of time dependent systems
  of differential equations (DEs) via the method of lines. COFFEE
  is primarily implemented in Python. It imposes very few requirements
  on users and was written with PEP8 \cite{pep8} as the guiding philosophy.
  Although COFFEE cannot compete with some
  existing numerical integrators for speed, it offers a low barrier for 
  use and substantial flexibility.

  COFFEE was specifically developed to compute solutions to
  a system of hyperbolic 
  partial differential equations (PDEs) 
  that represent Friedrich's conformal field equations
  \cite{friedrich1995einstein}. It
  has been used in eight research projects to numerically study
  the conformal properties of general relativity,
  \cite{beyer2012numerical,doulis2013second, beyer2014linearized, beyer2014spin,
    beyer2014numerical,frauendiener2014numerical, doulis2017global, beyer2017numerical}.
  As an illustration of the capabilities
  of COFFEE, in \cite{beyer2017numerical} it
  was used to solve a system of PDEs in the form of an Initial Boundary Value
  Problem (IBVP) containing 46 variables and 45 constraints on two different
  high performance clusters using up to 200 processes. It evolved the
  system in time for a range of resolutions, approximated spatial derivatives
  in a number of ways, stably imposed user-given boundary conditions
  and stored the data in HDF files. 
  Post-processing scripts demonstrated convergence and stability of
  the computed solution and produced
  visualizations of the output. 
  COFFEE contains the tools
  necessary to rigorously investigate the numerical evolution of a
  system of time dependent PDEs.

  COFFEE is unique in that, there is no
  computational PDE software
  designed with the philosophy of user-friendliness and flexibility 
  that has been used to solve complex and
  challenging systems of equations, for example those derived from Einstein's field equations.
  This contrasts COFFEE with existing software like
  Cactus \cite{goodale2002cactus}, Chombo \cite{colella2009chombo} or 
  PETSc \cite{balay2004petsc}. 

\section{Software description}
  \label{section_software_description}


  COFFEE was designed to 
  significantly reduce the amount of work needed to write code to
  solve systems of equations.
  Thus, despite its numerical nature, COFFEE is implemented in Python
  and relies heavily on \python{numpy}, \python{mpi4py}, 
  \python{hdf5} and custom \cinline{C} code.
  Implementation in Python
  has obvious disadvantages. For example; as Python is an interpreted language
  syntax can have a large impact on speed of execution (compare for loops to
  list comprehensions), there are structural issues with interpreted
  languages (in the case of Python this is the reason for the GIL),
  additional overhead in the translation of Python script to Python byte code
  and then to machine instructions, and a lack of
  the compile time checks that are found in strongly typed languages.
  Nevertheless the papers
  \cite{beyer2012numerical,doulis2013second, beyer2014linearized, beyer2014spin,
    beyer2014numerical,frauendiener2014numerical, doulis2017global, beyer2017numerical}
  demonstrate that COFFEE is capable of solving technically challenging
  and computationally intensive systems of PDEs.

  Implementation in Python also has advantages. Of particular note is duck typing,
  dynamic introspection and code injection, 
  which reduces the need for the user to 
  conform to strict programming patterns and 
  understand the ``COFFEE way of doing things''.
  Conforming to the philosophy of Python, COFFEE evaluates code 
  as given and fails fast, i.e. stops the simulation rather than continuing with a potential flaw. Before each iteration of the simulation,
  a collection of ``actions'' are run. Each action is an arbitrary piece of
  user code
  that has complete access to all data at the current time step and
  almost
  all objects performing the simulation. 
  This gives users substantial complete control over the simulation.
  
  The COFFEE code base
  has been written for readability over performance (except for code
  dealing with the MPI and HDF API's). The code contains plenty of comments
  highlighting trickier portions of code, why certain algorithms were chosen
  and portions of code that are bug prone under change.
  As a general point, we believe that code should not be
  viewed as an immutable body of work but rather like working on a whiteboard:
  added to and altered as needed.
  Hence, we expect users to directly alter
  COFFEE's code based whenever more convenient than other methods of
  changing the simulation; e.g.\ run time control can be exercised
  in actions or in system objects described in Section
  \ref{section_software_description}.
  Due to the technicalities of working with MPI
  and HDF API's, however, caution should be exercised
  when editing the
  \python{mpi}, \python{actions.hdf_output} and \python{io.simulation_data}
  modules.

  The core functionality of COFFEE is an MPI-enabled
  implementation of the method of lines with code to
  support spectral and finite difference techniques for spatial derivatives
  over clusters of computers.
  Thus it is an implementation of
  the numerical methods required for evolving time dependent
  systems of ODEs and PDEs,
  e.g.\ parabolic and hyperbolic systems. 
  Of particular note,
  COFFEE includes code for the simultaneous approximation
  (SAT) method \cite{carpenter1994time} for
  imposing stable boundary conditions, the papers
  \cite{carpenter1999stable, diener2007optimized, strand1994summation}
  for
  summation-by-parts finite difference operators, and
  \cite{huffenberger2010fast} for fast spin-weighted spherical harmonic
  transforms for numerical implementation of the $\eth$-calculus (eth-calculus),
  see for example \cite{penrose1987spinors}.
  IO uses HDF5 for data storage.
  It has been run on desktop workstations, 
  on a cluster of computers at the University
  of Otago and on
  the New Zealand eScience Infrastructure's high performance
  computing cluster.

  \begin{landscape}
    \begin{figure}[!ht]
      \begin{center}
        \begin{tikzpicture}
          \begin{umlseqdiag}
            \umlobject[no ddots]{ibvp}
            \umlobject[no ddots]{solver}
            \umlobject[no ddots]{system}
            \umlobject[no ddots]{diffop}
            \umlobject[no ddots]{timeslice}
            \umlmulti[no ddots]{action}
            \begin{umlcall}[dt=6,padding=3,op={\python{initial_data(t, grid)}}, return={\python{tslice}}]{ibvp}{system}
            \end{umlcall}
            \begin{umlfragment}[type=loop,name=main]
              \begin{umlcall}[dt=6, op={\python{action(i, tslice)}}]{ibvp}{action}
              \end{umlcall}
              \begin{umlcall}[dt=6,op={\python{advance(t, tslice)}},return={\python{tslice}}]{ibvp}{solver}
                \begin{umlfragment}[type=loop, name=loop]
                  \begin{umlcall}[op={\python{evaluate(t, tslice)}},return={\python{tslice}}]{solver}{system}
                    \begin{umlcall}[padding=3,op={\python{diffop(tslice.data)}},return={\python{tslice}}]{system}{diffop}
                    \end{umlcall}
                    \begin{umlcall}[dt=6,op={\python{communicate()}}, with return]{system}{timeslice}
                    \end{umlcall}
                    \begin{umlcall}[dt=6,op={\python{boundary_slices()}}, with return]{system}{timeslice}
                    \end{umlcall}
                  \end{umlcall}
                \end{umlfragment}
              \end{umlcall}
            \end{umlfragment}
            \umlsdnode[dt=43.5]{action}
            \umlsdnode[dt=7.5]{timeslice}
            \umlsdnode[dt=24.5]{diffop}
            \umlsdnode[dt=5.5]{system}
            \umlsdnode[dt=0.5]{solver}
            \umlnote[x=20, y=-10]{loop}{The number of times 
              this code block is executed 
              depends on the choice of solver.}
            \umlnote[x=20,y=-6]{main}{The outer loop 
              will be executed a sufficient number of
              times to perform the simulation.}
          \end{umlseqdiag}
        \end{tikzpicture}
      \end{center}
      \caption{A UML sequence diagram of COFFEE. The names of function calls
      and variables have been preserved in the diagram, \python{t} is the current
      time, \python{tslice} is a time slice of function data (passed in
      the \python{advance()} and \python{evaluate()} methods and
      created in the \python{initial_data()} call) or derivates of
      function data (returned in the \python{diffop()} call)
      or values of functions at intermediate times (returned from the
      \python{evaluate()} call), \python{i} is the number of the current
      iteration, \python{tslice.data} is the values of the function at the
      current time.  Note that the lifetime of the \python{tslice} object
      is strictly speaking incorrect as different time slices are used in
      each of the inner fragment loops. The \python{Solver} is responsible
      for creation of new time slices between each fragment evaluation. We
      have left this off the diagram as it only serves to complicate it.}
      \label{figure:data_flow}
    \end{figure}
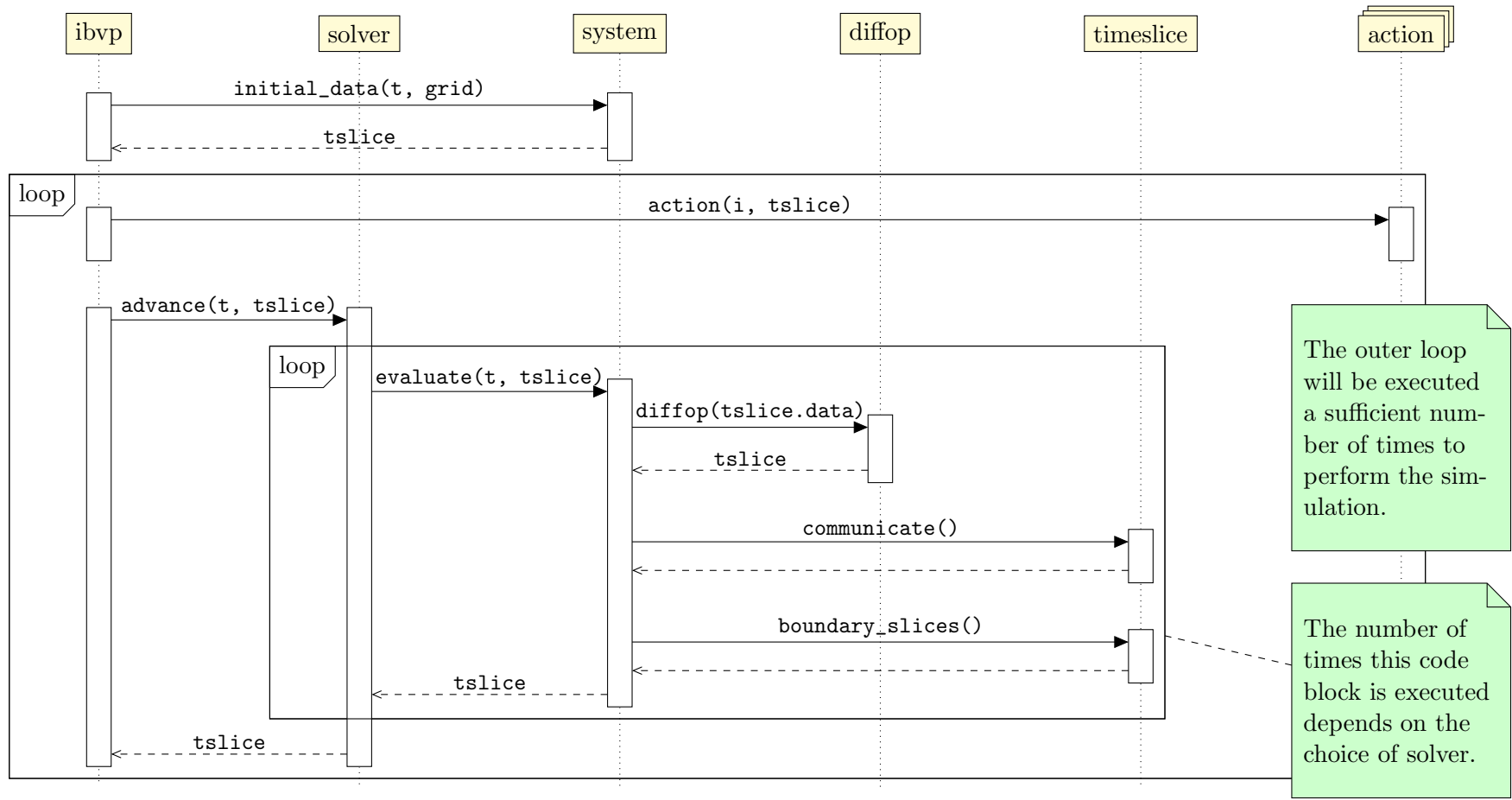
  \end{landscape}

  For the rest of this section, it may be helpful for the reader to refer to the UML sequence diagram of COFFEE, as shown in Figure \ref{figure:data_flow}.

  The \python{IBVP} class represents an initial boundary value problem. 
  To initialize this class objects behaving
  similarly to a ``Solver'', a \python{System} object, and
  a \python{Grid} object must be provided. The solver describes how steps
  along the lines of the simulation are performed. The system calculates
  the time derivative of the functions being simulated.
  The grid object describes the domain of the functions and manages the MPI API.
  Additional options, which have sensible defaults,
  at initialization are a list of actions to be performed
  during simulation (e.g.\ data reduction, visualization, error calculation, 
  and so on), the maximum number of iterations and a minimum time step.
  
  Solvers are objects that know
  how to integrate one dimensional ODEs. An abstract base class is
  provided along with implementations of the Euler (explicit and implicit), the 4th order 
  Runge Kutta method and
  a variation of the 4th order Runge Kutta method that incorporates
  boundary data for intermediate steps. An adaptive 4th order Runge Kutta
  method has been implemented but not tested sufficiently for this release of
  COFFEE. The code for this will eventually be included in the repository.

  System objects represent the system
  of differential equations to be
  solved using the method of lines.
  An abstract base class is provided as a form of documentation of the otherwise
  implicitly assumed API provided by custom classes filling the role of a 
  \python{system} object.
  System classes must have a method that returns what the next timestep
  is to be (this allows for adaptive simulation), 
  a method (\python{evaluate()}) that
  calculates the time derivative at a particular point in time, 
  and methods that give initial and boundary values.

  To make life easier for the user a number of numeric (spatial) derivative operations
  have been implemented. Instances of these have been used in the 
  \python{evaluate()} method in published papers. 
  Code for the following operators is provided: 
  \begin{itemize}
    \item 11 finite difference stencils,
    \item 9 different implementations using the fast Fourier transform,
    \item The Geroch-Held-Penrose operators $\eth$ and $\eth'$,
    \item 7 summation by parts finite difference operators with 3 supporting
      dissipation operators.
  \end{itemize}
  The Geroch-Held-Penrose operators are supported by a module, \python{swsh}, which
  can calculate and manipulate spin-weighted spherical harmonics.

  The \python{Grid} object represents the domain of the functions being calculated.
  To instantiate a grid object the number of data points in each dimension,
  the bounds for the values of coordinates on the data points,
  an \python{MPIInterface} object, and
  an object describing the ``boundary data'' must be provided.
  The \python{MPIInterface} object 
  wraps 
  an instance of 
  \mintinline{C}{MPI_COMM} itself
  wrapped by the \python{mpi4py} module 
  (COFFEE has been tested against both MPICH and Open MPI).
  COFFEE \python{MPIInterface} objects understand how to 
  communicate sufficient data to neighbouring processes to allow
  simulation to continue when simulation is performed over multiple processes
  via MPI. 

  The abstract base class for boundary data, \python{ABCBoundary}, represents
  the information needed to determine what data has to be sent between sub-grids
  on different processes via a subclass of \python{MPIInterface}.
  The \python{ABCBoundary} objects
  draw a distinction between
  sub-grid ``edges'' that are internal and external to the full grid
  as represented by the \python{Grid} object. Each internal and external
  edge can, in principle, have differing numbers of ghost points and points
  on the ``boundary'' region that will be communicated. This data is
  communicate via the \python{boundary_slices()} method.

  Once simulation is started
  COFFEE enters a ``main'' loop contained in the \python{ibvp} class.
  First, the next
  timestep is determined. Second, each action in the action list
  is performed. Third, the values of the function at the current time
  plus the time step are calculated. 
  Fourth, the process is repeated until an exception is
  raised, e.g.\ an overflow occurs, or the final time is reached.

  All data in each iteration of the main loop is stored in a time
  slice object. Each time slice contains the function data, the grid,
  and the current time. 


  The last major portion of COFFEE worth discussing is actions. An action
  is a piece of code that does something with a timeslice. Actions are 
  user definable and can contain references to any object available at runtime.
  When called an action is passed the current timeslice.
  Actions have complete freedom and therefore
  are able to dynamically affect the
  simulation. An action should subclass the \python{actions.Prototype} class
  and at least implement the function \python{_doit()}.


  COFFEE is supported by additional scripts that work on the resulting hdf files
  to produce the normal array of secondary derived information, such as 
  the calculation of
  convergence rates, errors, visualization, and manipulation of 
  data in hdf files.


  There are only a few requirements that need to be met
  before COFFEE can be used to compute solutions, see Section
  \ref{section_minimal_example} for an example or refer to one of the
  more detailed examples provided in the COFFEE repository.
  Users must provide:
  \begin{enumerate}
    \item an object with a method that returns the next time step $\Delta t$
      and a method that computes the time derivative of the system at a given
      point in time (which we call ``the system''),
    \item select an object which can solve ordinary
      differential equations (ODEs) or provide their own 
      (``the solver''), and
    \item specify the domain of the solution to be computed and what
      discretization to use (``the grid'').
  \end{enumerate}
  Once these components have been selected from libraries within COFFEE,
  or have been custom written, they are passed as arguments during object
  initialization to the \python{ibvp} class. The simulation can now be
  run by calling the \python{ibvp.run()} method.
  There are an array of additional options
  that can be specified involving IO, MPI topology and data communication,
  methods for calculation of spatial differences, length of simulation
  and forced evaluation at specific times, and real time visualization.

  To run a simulation the user must instantiate objects that behave like the
  COFFEE provided \python{System}, \python{Grid}, \python{Solver} and
  \python{IBVP} classes. COFFEE has existing implementations of
  \python{Grid}, \python{Solver} and \python{IBVP} classes that are
  sufficient for most simulations. Since the \python{System} class
  represents the system of DEs to be solved this is left to the
  user for implementation.
  Collectively these classes contain all
  information needed to perform a simulation. 
  In the research projects cited above initialization of these classes
  and the start of simulation has been collected in what we call
  a setup file. See Listing \ref{code_setup} for a minimal
  example of a setup file. The COFFEE repository contains more
  detailed examples.
  For use in research, the Otago numerical
  relativity group included 
  a plethora of runtime command line customizations in setup files, but
  this is not required.
  Typical options in a setup file involve:
  logging, differential operators, solvers, output settings,
  ``action lists'' and settings related to real time generation of
  visualizations.

  \subsection{Software Architecture} 


  In keeping with Python's philosophy COFFEE's architecture is flat,
  except where interaction with MPI or HDF API's is needed.
  As a consequence explicit code dependencies are either obvious,
  e.g.\ the IO system relies on \python{h5py}, or
  explicitly stated, e.g.\ \python{Grid} objects use instances
  of the \python{ABCBoundary} class which is in the
  \python{grid.grid} module along with the \python{Grid} class.
  In order to document the implicit dependencies which result
  from Python's reliance on duck typing numerous abstract base classes
  are provided. These classes document the otherwise implicit assumptions
  made about class API's. We encourage users to subclass abstract base
  classes, though this is not required.

  Since the implicit dependencies are the most likely to cause issues for
  new users we briefly describe them. 
  Figure \ref{figure:data_flow}
  describes the expected flow of data during
  initialization of the data and
  one iteration of the simulation and therefore presents a
  schematic view of the implicit dependencies internal to COFFEE.

\section{Illustrative Examples}\label{section_minimal_example}

  We give an example of code that solves the one dimensional wave equation
  in the code Listings \ref{code_system} and \ref{code_setup}.
  Code from these listings can be found in the COFFEE repository. 
  The first file,
  given in Listing \ref{code_system} defines the
  system object. It specifies what spatial differential operator to use,
  how to calculate a time step, the initial data and how to calculate
  the time derivatives of the solution.
  The second file, given in Listing \ref{code_setup}, initializes
  the objects necessary for simulation and hands them to an \python{IBVP}
  object which manages the main simulation loop.

  \begin{listing}
    \begin{minted}{python}
      # file name: OneDwave.py
      import numpy as np

      from coffee.tslices import tslices
      from coffee.system import System
      from coffee.diffop.sbp import sbp

      class OneDwave(System):

          def __init__(self):
              self.D = sbp.D43_2_CNG()

          def timestep(self, u):
              return 0.4 * u.domain.step_sizes[0]
              
          def initial_data(self, t0, grid):
              axis = grid.axes[0]
              rv = 0.5 * np.exp(-10 * (axis - axis[int(axis.shape[0] / 2)])**2)
              return tslices.TimeSlice([rv, np.zeros_like(rv)], grid, t0)
          
          def evaluate(self, t, Psi):
              f0, Dtf0 = Psi.data
              DxDxf = np.real(self.D(f0, Psi.domain.step_sizes[0]))
              DtDtf = DxDxf
              DtDtf[-1] = 0.0
              DtDtf[0] =  0.0
              return tslices.TimeSlice([Dtf0, DtDtf], Psi.domain, time=t)
    \end{minted}
    \captionof{listing}{A code listing for the file that defines
      the \protect\python{system} 
      object representing the one dimensional wave equation
      with Dirichlet boundary conditions.}
    \label{code_system}
  \end{listing}

  \begin{listing}
    \begin{minted}{python}
      # file name: OneDwave_setup.py
      from coffee import ibvp, solvers, grid
      from coffee.actions import gp_plotter

      from OneDwave import OneDwave

      system = OneDwave()
      solver = solvers.RungeKutta4(system)
      grid = grid.UniformCart((200,), [(0, 4)])
      plotter = gp_plotter.Plotter1D(
        system, 
        'set yrange [-1:1]', 
        'set style data lines'
      )

      problem = ibvp.IBVP(solver, system, grid=grid, action=[plotter])
      problem.run(0, 50)
    \end{minted}
    \captionof{listing}{The listing for the file that
      initiates simulation. Objects representing the
      system, a solver and a grid over which
      the solution will be calculated are initialized.
      The grid is chosen to have 200 steps over the interval
      $(0,4)$. The simulation runs over the time interval $(0,50)$. 
      The \protect\python{ibvp} object plots the output using gnuplot.
    }
    \label{code_setup}
  \end{listing}

  \begin{figure}[ht]
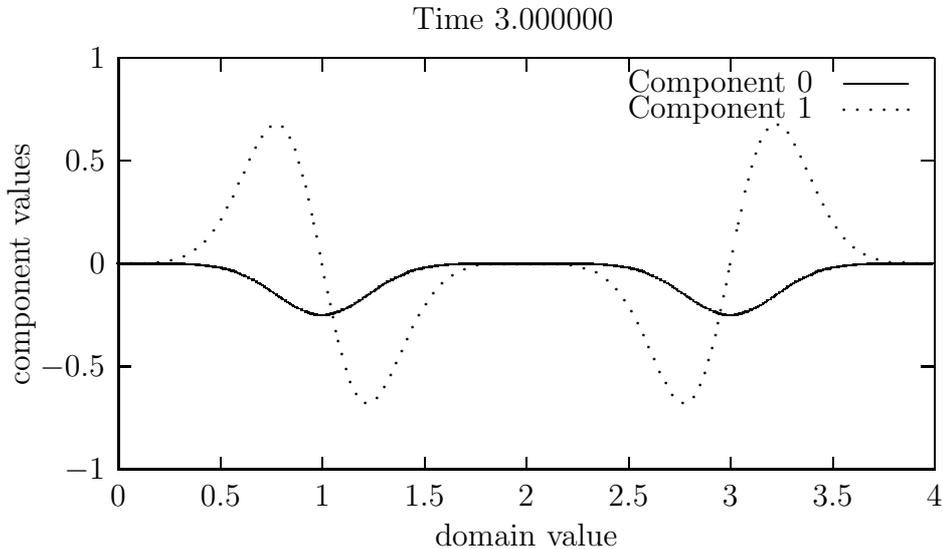

    \include{OneDWave}
    \caption{A graph of the function computed by Listings
      {\ref{code_system} and \ref{code_setup}}. Component $0$ is the value
      of the function that solves the wave equation specified in
      Listing {\ref{code_system}}, while component 1 is its derivative.
      The time has been chosen to show the function after the first reflection
      off both boundaries but before the separate wave packets merge for the
      first time. You can replicate this graph by using the 
      \protect\python{gpl_plotter} action to run only at time 3.0.}
    \label{code_figure}
  \end{figure}

\section{Impact}
  \label{sec:impact}



  There are only a few other packages currently available that incorporate a
  similar set of features to COFFEE.  Each of these has different philosophies,
  structure and goals. For example, the Cactus code
  \cite{goodale2002cactus,cactuscode} is a very large, community-driven project
  developed over many years. However the philosophy of Cactus
  and that of COFFEE differ greatly;
  Cactus is extremely optimized, written mostly in Fortran and C, and is very
  strict on how things are done and what the user can do. COFFEE on the other
  hand is written in Python and is designed for ease-of-use and flexibility.
  These properties make COFFEE and Cactus natural complements of each other and
  useful in their own right. 
  In particular, COFFEE, although powerful enough to
  satisfy advanced programmers, is also aimed at users from a variety of
  different fields that may have limited programming skills. 
  These users
  want a simple way to numerically evolve systems of ODEs or PDEs and are not
  necessarily worried about speed. 

  To help reduce runtime, COFFEE is MPI-parallelized: The computational domain is split into smaller domains, each having its own memory and dedicated core. The performance increase using MPI is exemplified with the strong scaling test results given in Table \ref{table:mpitest}. It is seen that increasing the number of MPI proceesses decreasing runtime up to around 32 cores, after which increasing MPI processes starts to increase computational time again. This is expected as the total gridsize is fixed, and the point at which increasing the number of MPI processes stops decreasing runtime will increase with an increased total gridsize. For a sufficiently large number of MPI processes, communication between each subgrid has increased to a point where the communication itself is now the bottleneck.

  User interaction with MPI in COFFEE is minimal, only a few lines of Python code are needed in the setup file, detailing: the dimension of the grid, the periodicity (if any), and the topology (e.g. Cartesian). A few more lines communicating data between processes in the system file may also be needed, e.g. for spatial derivative approximations. Examples of how to do the above are given in COFFEE's repository, see Table \ref{table:Codemetadata}.

  \begin{table}[!h]
    \begin{tabular}{|l|l|l|l|l|l|l|l|l|}
    \hline
    \textbf{\# of processes} & 1 & 2 & 4 & 8 & 16 & 32 & 64 & 128 \\
    \hline
    \textbf{Runtime (seconds)} & 256.6 & 190.4 & 117.7 & 71.9 & 61.3 & 54.5 & 54.8 & 58.4 \\
    \hline
    \end{tabular}
    \caption{A strong scaling test, using the one-dimensional system of symmetric
    hyperbolic PDEs described in \cite{frauendiener2014numerical}, with: 12801
    equi-distant spatial gridpoints between $z=-1$ and $z=+1$, spatial stepsize
    $\Delta z$, CFL of 0.5, temporal stepsize of $CFL*\Delta z$, spatial
    derivatives approximated with a fourth-order finite differencing operator
    with the summation-by-parts property \cite{strand1994summation} and boundary
    conditions implemented with the SAT method \cite{carpenter1994time}. The
    simulations were run on New Zealand eScience Infrastructure's Mahuika cluster
    which has 8,136 cores in 226 $\times$ Broadwell (E5-2695v4, 2.1 GHz, dual socket 18
    cores per socket) compute nodes.}
    \label{table:mpitest}
  \end{table}

  Further, COFFEE contains an implementation of spin-weighted spherical
  harmonics using the optimized transform
  algorithm of \cite{huffenberger2010fast}. 
  Current research involving COFFEE involves a modified version of \cite{huffenberger2010fast} which is 
  optimized for axi-symmetry. 
  This code is not released in this version of COFFEE but will be included
  after sufficient testing has been completed.
  This specialized spin-weighted spherical harmonic code was used in 
  \cite{beyer2016numerical} and \cite{beyer2017numerical}.




  The COFFEE package is extremely versatile as it presupposes very little about the
  system that the user inputs. Thus if a process can be modelled over time by a
  differential equation or a system of differential equations then it can be
  numerically evolved in COFFEE. Of course whether or not a numerical solution
  can be found will depend on the specifics of the system of equations and
  the chosen numerical methods.



\section{Conclusions}


  COFFEE is a user-friendly Python package for the numerical evolution of (a
  system of) ODEs and PDEs. It contains a wide variety of numerical algorithms
  for marching in time, approximating spatial derivatives and stably imposing
  boundary conditions as well as being MPI-parallelized by the splitting of the
  computational domain. It has been rigorously tested during multiple research
  projects and has functionality through the \emph{actions class} for
  performing user-defined tasks during the evolution. COFFEE is ideally suited
  to users that hold user-friendliness above absolute speed and want
  flexibility to taylor the code to their particular problem.

\section*{Acknowledgements}
The authors wish to acknowledge the use of New Zealand eScience Infrastructure (NeSI) high performance computing facilities and consulting support as part of this research. New Zealand's national facilities are provided by NeSI and funded jointly by NeSI's collaborator institutions and through the Ministry of Business, Innovation \& Employment's Research Infrastructure programme. URL \url{https://www.nesi.org.nz}.





\bibliographystyle{elsarticle-num} 
\bibliography{coffee}






\section*{Required Metadata}

\section*{Current code version}


\begin{table}[!h]
\begin{tabular}{|l|p{6.5cm}|p{6.5cm}|}
\hline
\textbf{Nr.} & \textbf{Code metadata description} & {} \\
\hline
C1 & Current code version & v1 \\
\hline
C2 & Permanent link to code/repository used for this code version & https://gitlab.com/thebarista/ $\;\;$coffee.git\\
\hline
C3 & Legal Code License   & GNU General Public License (GPL) \\
\hline
C4 & Code versioning system used & git \\
\hline
C5 & Software code languages, tools, and services used & Python 2.7, C, MPI, HDF5, numpy, spinsfast, libfftw3 \\
\hline
C6 & Compilation requirements, operating environments & Dependencies: \python{mpi4py} with
  compatible MPI implementation, \python{h5py} with compatible HDF implementation, \python{numpy}, \python{scipy}, \python{Gnuplot}, \python{matplotlib} and a variety of standard Python modules (e.g. \python{abc}, \python{logging} and \python{math}). \\
\hline
C7 & If available Link to developer documentation/manual & documentation provided in the code and the repository \\
\hline
C8 & Support email for questions & maintainers are contactable via the gitlab repository\\
\hline
\end{tabular}
\caption{Code metadata (mandatory)}
\label{table:Codemetadata}
\end{table}




\end{document}